\documentclass[%
 reprint,
nobibnotes,
bibnotes,
 amsmath,amssymb,
 aps,
]{revtex4-1}
\usepackage{MnSymbol}
\usepackage{natbib} 
\usepackage{xcolor}
\usepackage{soul}
\usepackage{hyperref}
\usepackage{wasysym} 
\usepackage{graphicx}
\usepackage{tikz}
\usetikzlibrary{calc}
\usepackage{dcolumn}
\usepackage{bm}

\tikzstyle{MyVertex}=[draw,circle,fill=black,inner sep=0,minimum size=3pt]
\tikzstyle{MyVacantVertex}=[draw,circle,fill=white,inner sep=0,minimum size=4pt]

\tikzstyle{MyDashedEdge}=[draw,dotted,line width=1pt]
\tikzstyle{MyFullEdge}=[line width=1pt]

\begin{document}

\preprint{Draft}


\title{An alternative expression of message passing on networks}

\author{Peter Mann}
\email{peter.mann@data-insights-ai.co.uk}
\affiliation{Data Insights AI, CodeBase Edinburgh, Argyle House, 3 Lady Lawson street, Edinburgh, EH3 9DR, United Kingdom}

\author{Simon Dobson}
\affiliation{School of Computer Science, University of St Andrews, St Andrews, Fife KY16 9SX, United Kingdom }

\date{\today}

\begin{abstract}
    Message passing techniques on networks encompasses a family of related methods that can be employed to ascertain many important properties of a network. It is widely considered to be the state of the art formulation for networked systems and advances in this method have a wide impact across multiple literatures. One property that message passing can yield is the size of the largest connected component in the network following bond percolation. In this paper, we introduce an alternative method of finding this value that differs from the standard approach. Like the canonical approach, our method is exact on trees and an approximation on arbitrary graphs. We show that our method lends itself to the description of a variety of generalisations of bond percolation such as sequential percolation and non-binary percolation and can yield information about the local environment of a node in percolation equilibrium that the traditional approach cannot.
\end{abstract}

\pacs{Valid PACS appear here}
\maketitle

\section{Introduction}

Bond percolation is a stochastic process in which the edges of a graph, $G(N,E)$, are \textit{occupied} with probability $\phi$, where $N$ is the number of nodes and $E$ is the number of edges in the graph. A cornerstone of statistical physics, bond percolation is a mathematical model that can be used to represent a range of phenomena from epidemics spreading through populations, fluid flow through porous media as well as the structural properties of crystalline materials.

Of primary interest to the bond percolation problem are the size of the percolating clusters (sets of nodes connected by occupied edges). In particular, the largest of these connected clusters - the giant connected component (GCC) displays phase properties as $\phi$ is varied. When $\phi$ is below some critical value $\phi_c$, the GCC contains $\mathcal O(1)$ nodes and the network is fractured into lots of disjoint components. As $\phi$ is increased past $\phi_c$, the GCC occupies a finite part of the graph. The fraction of the network occupied by the GCC, $S(\phi)$, is the order parameter of the percolation model, with $S(0)=0$ and $S(1)=1$.

Message passing is an analytical technique originating from statistical physics that has seen widespread use throughout computer science since the 1980s across a range of applications \cite{PhysRevLett.113.208702, PhysRevE.82.016101,Newman_2023, Mezard_Parisi_Zecchina_2002, Dorogovtsev_F._2022,Newman_2019}. Also known as belief propagation, message passing generalises transfer matrix method, the forward-backward algorithm, Kalman filtering and various peeling algorithms. It allows us to answer the question of whether a particular node, $i$, belongs to the GCC or not; and to find the size of the GCC itself, $S(\phi)$. The answer is exact when the graph is a tree and is otherwise an (often very good) approximate answer. For now we will assume that our graph is in fact a tree. Due to its ubiquity across several domains, advances in the fundamental techniques of the message passing algorithm have a broad range of applications to those scientific fields that consume it, such as computer science, artificial intelligence, deep learning, network science and statistical physics. In particular various neural network techniques such as Graph Neural Networks (GNNs) and Graph Convolution Networks (GCNs) use message passing to exchange messages between neighbours in order to update the feature representations of their nodes.

Let $\mathcal N(i)$ be the set of neighbours of node $i$. The probability, $\mu_i(\phi)$, that node $i$ does not belong to the GCC is a function of the probabilities that its neighbours $j\in \mathcal N(i)$, fail to connect $i$ to the GCC when $i$ is removed from the graph. The function can be written as the product that each edge $j$ fails to lead to the GCC. In turn, this is the sum of the probabilities of the different ways that $j$ might fail to connect $i$ during the percolation process. We can consider the possible ways that an edge between $i$ and $j$ might not lead to the GCC. Firstly, the edge could be unoccupied with probability $1-\phi$ and so not connect to $i$ in the first place; or, it could be occupied, but $j$ itself not be connected to the GCC, $\phi\mu_{i\leftarrow j}$. The probability $\mu_i$ is then given as the product that each neighbour fails
\begin{equation}
    \mu_i = \prod_{j\in \mathcal N(i)} (1-\phi + \phi\mu_{i\leftarrow j})\label{eq:mu_i}.
\end{equation}
The removal of $i$ is important as it ensures that the neighbours are uncorrelated to one another (assuming $G$ is a tree). This means that Eq \ref{eq:mu_i} \textit{can} in fact be written as the product of edges considered in isolation, over index $j$. If one of the neighbours did still belong to the GCC despite node $i$s removal, node $i$ must not have been its source of connection to the GCC. The graph without node $i$ is called the \textit{cavity graph} of $G$ at $i$, denoted $G(i)$. If $G(N,E)$ was not a tree, then two neighbours of $i$, $j_1,j_2\in\mathcal N(i)$, might still be connected in the cavity graph of $i$ by some longer route, and this would mean that the probabilities $\mu_{i\leftarrow j_1} $ and $\mu_{i\leftarrow j_2}$ were correlated and the subsequent analysis, including  \ref{eq:mu_i}, would be inexact. 

To use Eq \ref{eq:mu_i} we must calculate each probability $\mu_{i\leftarrow j}$. The probability that $j$ is not in the giant component in the cavity graph at $i$ is equal to the probability that none of $j$’s neighbors, other than $i$, belong to the GCC
\begin{equation}
    \mu_{i\leftarrow j} = \prod_{k\in \mathcal N(j)\backslash i} (1-\phi+\phi\mu_{j\leftarrow k}),\label{eq:mu_j}
\end{equation}
where ${k\in \mathcal N(j)\backslash i}$ is the set of neighbours of $j$ apart from node $i$. There is an equation of the form of Eq \ref{eq:mu_j} for each directed edge pair in the network, totalling a system of $2E$ expressions. This system of equations is known as the message passing equations and can be solved by numerical iteration to its fixed point. Once each $\mu_{i\leftarrow j}$ has been solved for all directed edge pairs in the graph, Eq \ref{eq:mu_i} can be evaluated for each node. In turn, these values can be used to find the size of the GCC
\begin{equation}
    S(\phi) = 1-\frac{1}{N}\sum_i\mu_i.\label{eq:S_canonical}
\end{equation}
For $\phi$ less than some critical value $\phi_c$, the system has a trivial solution of $S=0$ and no GCC exists. For $\phi\geq \phi_c$ the solution space bifurcates and an additional fixed point of $\mu_i<1$ appears, indicating that a GCC exists in the graph. Eq \ref{eq:S_canonical} relies on the mutual exclusivity of the binary state equilibrium of the percolation process and for the purpose of this paper we refer to Eqs \ref{eq:mu_i}, \ref{eq:mu_j} and \ref{eq:S_canonical} as the \textit{canonical approach} to message passing. 

In this paper, we introduce an alternative expression of $S(\phi)$ by considering a different family of $2E$ messages $\nu_i$ over the edge set. The collective knowledge of $4E$ variables can then be used to introduce a third reformulation for $S(\phi)$ in terms of both $\mu$ and $\nu$ message sets. In this new re-formulation we gain insight into the local neighbourhood of vertices embedded into the GCC that is not present with the canonical approach. This is because we can examine piecewise the contribution of each edge configuration to the percolation equilibrium. We do not claim that our method is \textit{easier} to apply than the canonical approach: in fact, it is more complicated and computationally slower. However, our re-formulation is particularly suitable when applied to processes that extend the ordinary percolation model. We explore this by considering \textit{sequential percolation} and \textit{non-binary percolation} processes using our approach.

\section{Theoretical}

\subsection{Reformulation of message passing}
\label{subsec:theoretical1}
The essence of our reformulation is as follows. In the canonical approach to message passing, we considered all possible combinations of edge occupations in which a neighbour failed to connect node $i$ to the GCC. In our reformulation, we enumerate all possible combinations that node $i$ \textit{is} successfully connected to the GCC by its set of neighbours. This is a non-trivial enumeration and we must consider the combinatorics of the neighbour set as a whole as the edges are no longer independent of one another. The probability of connection $\nu$ can be treated as an alternative message set over $G$. Knowledge of both $\mu$ and $\nu$ message sets gives us complete knowledge over the percolating system, with detail that is not found in the canonical approach.

\begin{figure}[ht!]
\centering
\includegraphics[width=0.345\textwidth]{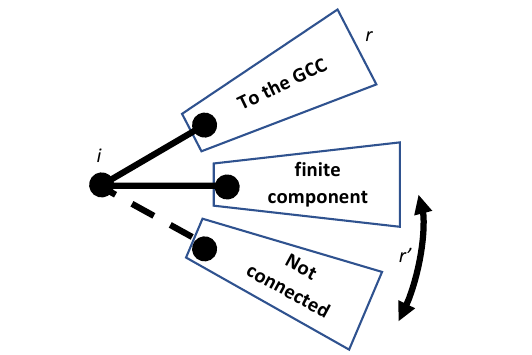}
\caption{A graphical representation of the enumeration. Solid edges are occupied whilst dashed edges are unoccupied.}
\label{fig:picture}
\end{figure}

To begin, let $\pi_{i}(s, \phi)$ be the probability that the sum of the number of nodes reachable along each edge from $i$ is $s-1$. For brevity, we will drop the explicit dependence on $\phi$ from our notation, $\pi_{i}(s, \phi)\equiv\pi_i(s)$. We can write this as
\begin{equation}
    \pi_i(s) = \sum_{\{s_j:j\in \mathcal N(i)\}}\left[\prod_{j\in \mathcal N(i)}\pi_{i\leftarrow j}(s_j)\right]\delta\left(s-1,\sum_{j\in\mathcal N(i)}s_j\right),\label{eq:pi_i}
\end{equation}
where $\pi_{i\leftarrow j}(s_j)$ is the probability that the edge $(i, j)$ leads to $s_j$ nodes, $\mathcal N(i)$ is the set of neighbours of $i$, $\delta(a,b)$ is the Kronecker delta and $\{s_j:j\in \mathcal N(i)\}$ is the set of all values of $s_j$ for each neighbour $j$ to node $i$. To calculate $\pi_i(s)$ we require a closed form expression for $\pi_{i\leftarrow j}(s_j)$. Let us distinguish three kinds of edges: i) edges that are occupied and lead to the GCC, ii) edges that are occupied and lead to a finite cluster in $G(i)$, and iii) edges that are not occupied, see Fig \ref{fig:picture}.

\begin{figure}[ht!]
\centering
\includegraphics[width=0.45\textwidth]{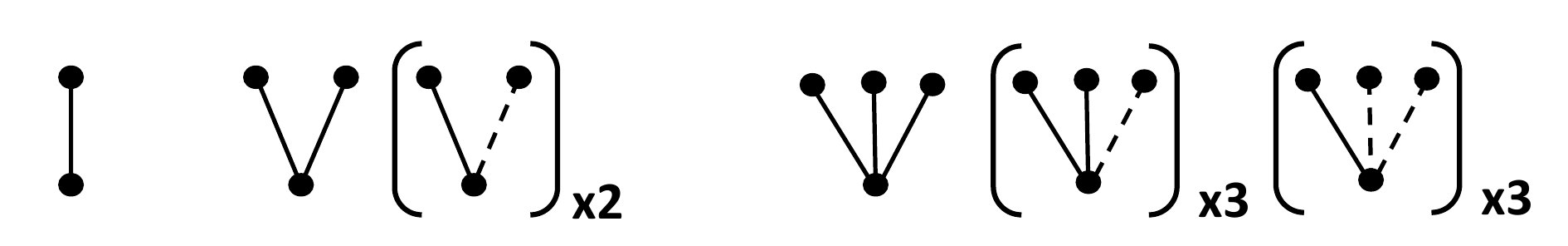}
\caption{A graphical representation of the first few terms of the polynomials that arise during the direct enumeration of all configurations that connect the focal node to the GCC. Occupied edges are solid lines whilst unoccupied edges are dashed; in each case the focal node $i$ is the bottom node. Configurations that are related by symmetry are enclosed in brackets subscripted by their degeneracy.}
\label{fig:inverse}
\end{figure}

Partitioning the neighbours $\mathcal N(i)$ into those that lead to the GCC, $X$, and those that do not, $\bar X$, we can write
\begin{equation}
    \prod_{j\in \mathcal N(i)}\pi_{i\leftarrow j}(s_j) = \prod_{r\in X}\pi_{i\leftarrow r}(s_r)\prod_{r'\in  \bar X}\pi_{i\leftarrow r'}(s_{r'}).
\end{equation}
The set $\bar X$ is the complement set of $X$, containing all members of $\mathcal N_i$ that are not in $X$. Variables $r$ and $r'$ index the members of $X$ and $\bar X$, respectively.
Substituting this into Eq \ref{eq:pi_i} we find
\begin{align}
    \pi_i(s) =& \sum_X\sum_{\{s_r:r\in X\}}\sum_{\{s_{r'}:r'\in \bar X\}}\left[\prod_{r\in X}\pi_{i\leftarrow r}(s_r)\prod_{r'\in \bar X}\pi_{i\leftarrow r'}(s_r')\right]\nonumber\\
    &\times\delta\left(s-1,\sum_{r\in X}s_r+\sum_{r'\in \bar X}s_r'\right),\label{eq:pi_i_mod}
\end{align}
where the summation over $X$ is over all non-empty subsets of $\mathcal N(i)$. The set $\{s_r:r\in X\}$ is the set of all possible values of $s_r$, and similarly for $\{s_{r'}:r'\in \bar X\}$. The generating function $\nu_i(z)$ for this expression is 
\begin{equation}
    \nu_i(z) = \sum_{s=1}^\infty \pi_i(s)z^s,
\end{equation}
which is the probability that node $i$ belongs to the GCC in the percolation equilibrium. Inserting the following logic
\begin{equation}
    z^s=zz^{s-1} = zz^{\sum_rs_r+\sum_{r'}s_{r'}},
\end{equation}
we have 
\begin{align}
    \nu_i(z) =& z\sum_X\left(\prod_{r\in X} \left[ \sum_{\{s_r:r\in X\}} \pi_{i\leftarrow r}(s_r)z^{\sum_{r}s_r} \right]\right. \nonumber\\
    &\left.\times\prod_{r'\in \bar X}\left[ \sum_{\{s_{r'}:{r'}\in\bar X\}} \pi_{i\leftarrow r'}(s_{r'})z^{\sum_{r'}s_{r'}} \right]\right).
\end{align}
The inner summation and the product can be exchanged due to the independence of each edge in the cavity graph technique \cite{Newman_Strogatz_Watts_2001}.
This quantity can be enumerated by accounting for all of the configurations of occupied edges that connect the focal node to the GCC, see Fig \ref{fig:inverse}.

Defining $\nu_{i\leftarrow r}(z)$ as the probability that neighbour $r$ is connected to the GCC
\begin{equation}
    \phi^{|X|}\nu_{i\leftarrow r} = \sum_{\{s_r:r\in X\}} \pi_{i\leftarrow r}(s_r)z^{\sum_{r}s_r}, \label{eq:leading}
\end{equation}
we finalise the expression for $\nu_i(z)$ as
\begin{equation}
    \nu_i(z)=z\sum_{l=1}^{|\mathcal N(i)|}\sum_{X\in Y_l}P_\phi(X)\label{eq:nuij},
\end{equation}
where $P_\phi(X)$ is the following probability
\begin{equation}
    P_\phi(X) = \left(\phi^{|X|}\prod_{r\in X}\nu_{i\leftarrow r}\prod_{r'\in \bar X}[1-\nu_{i\leftarrow r'}\phi]\right),
\end{equation}
and where $Y_l$ is the set of all subsets of length $l$ of the set of neighbours $\mathcal N(i)$ without replacement; set $X$ belongs to $Y_l$, $|X|$ is the cardinality of $X$ and similarly for $|\mathcal N(i)|$. Note the leading factor of $\phi^{|X|}$ in Eq \ref{eq:leading} is because there must be exactly $|X|$ occupied edges for a given configuration of set $X$ over $\mathcal N(i)$. The probability $P_\phi(X)$ is simply the product of probabilities that a subset $X$ of the neighbours in $G(i)$ connects to the GCC and a subset $\bar X$ does not. The summation over $X\in Y_l$ simply accounts for the different configurations that the subset of size $l$ could be arranged in.

The probabilities $\nu_{i\leftarrow r}$ are the probabilities that neighbour $r$ connects $i$ to the GCC, which can be solved by writing a $2E$-dimensional message passing system of equations of the form
\begin{equation}
    \nu_{i\leftarrow r}(z) = z\sum_{l=1}^{|\mathcal N(r)\backslash i|}\sum_{T\in W_l\backslash i}\left( \phi^{|T|}\prod_{t\in T}\nu_{r\leftarrow t}\prod_{t'\in \bar T}[1-\nu_{r\leftarrow t'}\phi]\right)\label{eq:nujk},
\end{equation}
where $W_l\backslash i$ is the set of subsets of length $l$ of the set of neighbours $\mathcal N(r)$ of node $r$ apart from node $i$ and $|T|$ is the cardinality of set $T$. As previously, $\bar T $ is the complement set of $T$, containing all nodes in $\mathcal N(r)\backslash i$ that do not belong to $T$. From these quantities, the expectation value for the size of the GCC is given by 
\begin{equation}
    S(\phi) = \frac 1 N \sum_i \nu_i(1),\label{eq:sfrominf}
\end{equation}
where $N$ is the number of nodes in the graph.

\begin{figure}[ht!]
\centering
\includegraphics[width=0.48\textwidth]{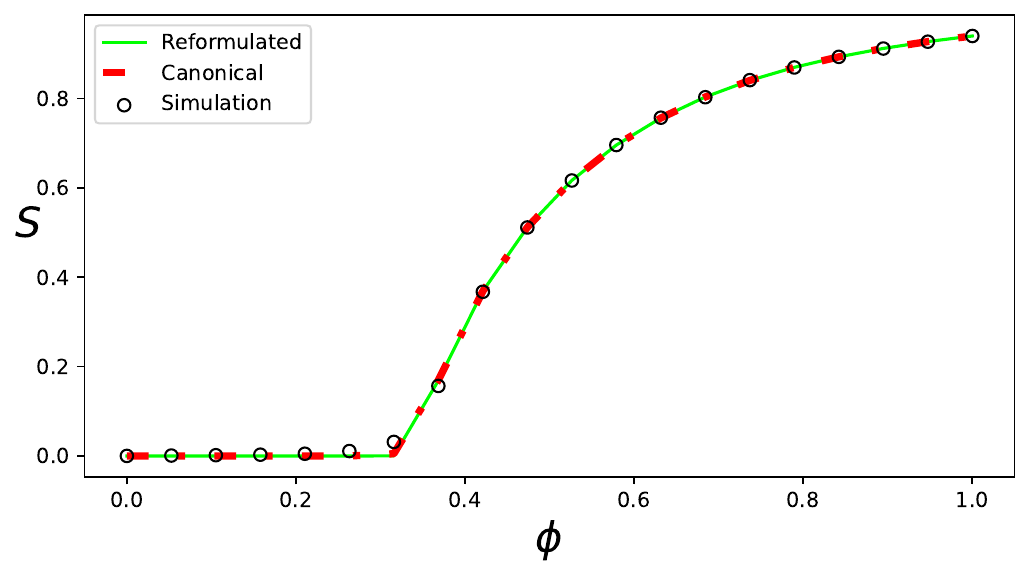}
\caption{A comparison of the calculated values of $S$ using both the canonical and reformulated message passing models alongside Monte Carlo simulations of bond percolations. The two methods predict the same numerical result to machine precision. The network is an Er\H{o}s-Renyi graph with 5000 nodes and an average degree of $3$ whilst simulations are the average of 100 repeats of percolation.}
\label{fig:compare}
\end{figure}

If we like, these new variables can be combined with the original set from Eq \ref{eq:mu_j} to re-write Eq \ref{eq:nujk} compactly in terms of $4E$ variables as 
\begin{equation}
    \nu_{i\leftarrow r}(z) = z\sum_{l=1}^{|\mathcal N(r)\backslash i|}\sum_{T\in W_l\backslash i}\left( \phi^{|T|}\prod_{t\in T}\nu_{r\leftarrow t}\prod_{t'\in \bar T}(1-\phi+\mu_{r\leftarrow t'}\phi)\right),\label{eq:nujkmu}
\end{equation}
although this substitution is not required to solve for $\nu_{i\leftarrow r}(z)$. With reference to Fig \ref{fig:picture}, $1-\phi$ accounts for the unoccupied edges, whilst $\mu_{r\leftarrow {t'}}\phi$ accounts for the edges that lead to finite clusters. Comparison of Eqs \ref{eq:sfrominf} and \ref{eq:S_canonical} yields the following relation between the two descriptions
\begin{equation}
    1 = \frac 1 N \sum_i\nu_i(\phi) + \frac 1N\sum_i\mu_i(\phi).\label{eq:big_res}
\end{equation}
This final relation subtly exhibits the mutual exclusivity of the two analytical descriptions of the percolation equilibrium and shows that we have complete knowledge of the system by accounting for all possible configurations of the equilibrium. We show an comparison of the two approaches alongside simulation in Fig \ref{fig:compare}.

Below the critical point, the probability that a node belongs to the GCC is zero, since a GCC does not exist in the network. The system described by Eq \ref{eq:nujk} has the trivial solution $0=0$. At the critical value of the control parameter, $\phi_c$, the solution space bifurcates and non-zero solutions can be found. By linearising Eq \ref{eq:nujk} around $\nu_{i\leftarrow r}\approx \epsilon_{i\leftarrow r}$ for some small parameter $\epsilon_{i\leftarrow r}$, the system at the critical point is given by 
\begin{equation}
\epsilon_{i\leftarrow r} \approx \sum_{l=1}^{|\mathcal N(r)\backslash i|} \sum_{T\in W_l\backslash i}\left(\phi_c^{|T|}\prod_t\epsilon_{r\leftarrow t}\right)\bigg|_{z=1}.
\end{equation}
Solving this expression numerically for $\phi$ yields the critical point $\phi_{c}$ and it is dominated by terms linear in $\epsilon_{i\leftarrow r}$. This means that configurations of occupied edges that have single occupancies have the highest measure associated around a GCC node at the critical point and therefore that the GCC is expected to be tree-like at this point.

The relation of this expression to the Hashimoto non-backtracking matrix and the implications for the community detection and centrality literature \cite{Krzakala_Moore_Mossel_Neeman_Sly_Zdeborova_Zhang_2013b, PhysRevE.90.052808}, remains unknown.

\subsection{The partition function \texorpdfstring{$\mathcal Z_{i,\phi}$}{Lg}}

The polynomial $\sum_{l,X} P_\phi(X)$ in Eq \ref{eq:nuij} is an important contribution to our understanding of the local environment of a node within the GCC of a percolation equilibrium. It can be regarded as a partition function $\nu_i(1)=\mathcal Z_{i, \phi}$ for the configurations of neighbours that also belong to the GCC with node $i$ and we will denote it as follows
\begin{equation}
    \mathcal Z_{i, \phi} = \sum_{l=1}^{|\mathcal N(i)|}\sum_{X\in Y_l}P_\phi(X)\label{eq:partition_function}.
\end{equation}
Here we will show that $\mathcal Z_{i, \phi}$ contains more information than can be obtained via the canonical approach and we will later use it to investigate the fine structure of the GCC surrounding a given node in the network. 

One of the primary benefits of Eq \ref{eq:nuij} over Eq \ref{eq:mu_i} is that it contains each possible configuration of edges that a node might experience in the percolation equilibrium. 
Immediately, this allows us to identify which edge configurations are associated with the highest measure and to better investigate occupation patterns across the network, tailored to each node and region in possibly heterogeneous environments. This yields information on the fine structure of a network beyond that which can be achieved using the canonical approach.

Whilst the sets $Y_l$ and $X$ from Sec \ref{subsec:theoretical1} can easily be enumerated computationally for arbitrary $k=|\mathcal N(i)|$, for small degrees they can be written by hand also. For example, when $k=1$ we have $P(\{j_1\}) = \phi\nu_{i\leftarrow j_1}$ for neighbour $j_1\in \mathcal N(i)$. For $k=2$ and $X\in [\{j_1\}, \{j_2\}, \{j_1, j_2\}]$ we have
\begin{align*}
    \mathcal Z_{i, \phi} =&\ \phi\nu_{i\leftarrow j_1}(1-\phi\nu_{i\leftarrow j_2})\nonumber\\
    &+ \phi\nu_{i\leftarrow j_2}(1-\phi\nu_{i\leftarrow j_1})  +\phi^2\nu_{i\leftarrow j_1}\nu_{i\leftarrow j_2}.
\end{align*}
For $k=3$ with neighbours $j_1,j_2,j_3\in \mathcal N(i)$ we have 
\begin{align*}
    X\in &[\{j_1\}, \{j_2\}, \{j_3\},\{j_1, j_2\}, \{j_1, j_3\}, \\& \{j_2, j_3\}, \{j_1, j_2, j_3\}],
\end{align*}
from which we obtain
\begin{align*}
    \mathcal Z_{i, \phi} =\ & \phi^3\nu_{i\leftarrow j_1}\nu_{i\leftarrow j_2}\nu_{i\leftarrow j_3} \nonumber\\
    &+ \phi^2\big[ \nu_{i\leftarrow j_1}\nu_{i\leftarrow j_2}(1-\nu_{i\leftarrow j_3}\phi)\nonumber\\
    & +\nu_{i\leftarrow j_1}(1-\nu_{i\leftarrow j_2}\phi)\nu_{i\leftarrow j_3}\nonumber\\
    &+ (1-\nu_{i\leftarrow j_1}\phi )\nu_{i\leftarrow j_2}\nu_{i\leftarrow j_3}\big]\nonumber\\
    &+ \phi\big[\nu_{i\leftarrow j_1}(1-\nu_{i\leftarrow j_2}\phi)(1-\nu_{i\leftarrow j_3}\phi) \nonumber\\
    &+ (1-\nu_{i\leftarrow j_1}\phi )\nu_{i\leftarrow j_2}(1-\nu_{i\leftarrow j_3}\phi) \nonumber\\
    &+ (1-\nu_{i\leftarrow j_1}\phi)(1-\nu_{i\leftarrow j_2}\phi)\nu_{i\leftarrow j_3}\big].
\end{align*}
It can be shown that these expressions are numerically equivalent to those obtained from the polynomial generated by the canonical approach from Eq \ref{eq:mu_i} by the following expression
\begin{equation}
    1 = \mathcal Z_{i, \phi} + \prod_j(1-\phi+\mu_{i\leftarrow j}\phi).\label{eq:Ztwo}
\end{equation}
However, whilst the polynomial in the canonical approach sums up the ways node $i$ fails to connect to the GCC, $\mathcal Z_{i, \phi}$ instead enumerates the configurations where node $i$ does connect. 

There are a number of different ways to extract useful information about the network from this polynomial and we will detail these next. Firstly, we can obtain the probability $\mathcal Q_{X'}(X)$ that a particular configuration $X'$ of neighbours appears around a given node $i$ in the GCC
\begin{equation}
    \mathcal Q_{i, X'}(X) = \frac{1}{\mathcal Z_{i, \phi}}P_\phi(X').
\end{equation}
This is a useful expression to examine the measure associated to each microstate that the focal node could be in. 

Next, we can find the probability that a particular node $j\in \mathcal N(i)$ belongs to the GCC along with node $i$ by summing over all configurations that contain $j$
\begin{equation}
    \mathcal R_{i,j}(X) = \frac{1}{\mathcal Z_{i,\phi}}\sum\limits_{X'}P_\phi(X'),
\end{equation}
with
\begin{equation}
    \{X'\in Y_l : j\in X';\  l\in 1,\dots,|\mathcal N(i)|\}.
\end{equation}
This can also be generalised to an arbitrary subset of $i$s neighbours
\begin{equation}
    \mathcal R_{i, j_1,\dots,j_n}(X) = \frac {1}{\mathcal Z_{i,\phi}}\sum\limits_{X'}P_\phi(X'),
\end{equation}
such that
\begin{equation}
    \{X'\in Y_l : (j_1,\dots,j_n)\in X';\ l\in n,\dots,|\mathcal N(i)|\}.
\end{equation}

Finally, we can also obtain the expectation value $\mathcal T_{i,k'}(X)$ that the focal node has degree $k'$ within the GCC by constraining the summation over $X'$ to those sets such that $|X'|=k'$
\begin{equation}
    \mathcal T_{i, k'}(X) = \frac{1}{\mathcal Z_{i,\phi}}\sum\limits_{X'}P_\phi(X'),\label{eq:Tki}
\end{equation}
and
\begin{equation}
     \{X'\in Y_{k'} :|X'|=k'\}.
\end{equation}
This is useful for investigating the structure of the giant component and can be thought of as the $\phi$-dependent degree distribution for a given node in the GCC. As with any distribution we have $ \sum_{k'}\mathcal T_{i,k'} = 1$.

We can use Eq \ref{eq:Tki} to compute the average degree for a given $\phi$ that a node will have in the GCC. For example in Fig \ref{fig:Tk_i_expr_stack} (top) we show how the components of $\mathcal T_{i,k'}$ change as a function of $\phi$ giving insight into the substructure of the GCC from the perspective of the local environment of the node within it. The average number of occupied edges of a node, $i$, in the GCC is given by the following expectation value
\begin{equation}
    \langle k_{i}'\rangle = \sum_{k'}k'\mathcal T_{i,k'}.\label{eq:degdist}
\end{equation}
We anticipate that the assortativity of the GCC can equally be investigated using this quantity in a similar, yet simpler fashion than \cite{Hasegawa_Mizutaka_2020}.

With all of these functions, $\mathcal Q_{i,X'}$, $\mathcal R_{i,j_1,\dots,j_n}$ and $\mathcal T_{i, k'}$ the important point to note is that we cannot obtain this information from the canonical approach. Whilst we can numerically obtain the value of $\mathcal Z_{i,\phi}$ by rearranging Eq \ref{eq:Ztwo}, it is only because we have access to the full polynomial from the right hand side of Eq \ref{eq:partition_function} that we can write the numerator of these quotients.

\begin{figure}[ht!]
\centering
\includegraphics[width=0.48\textwidth]{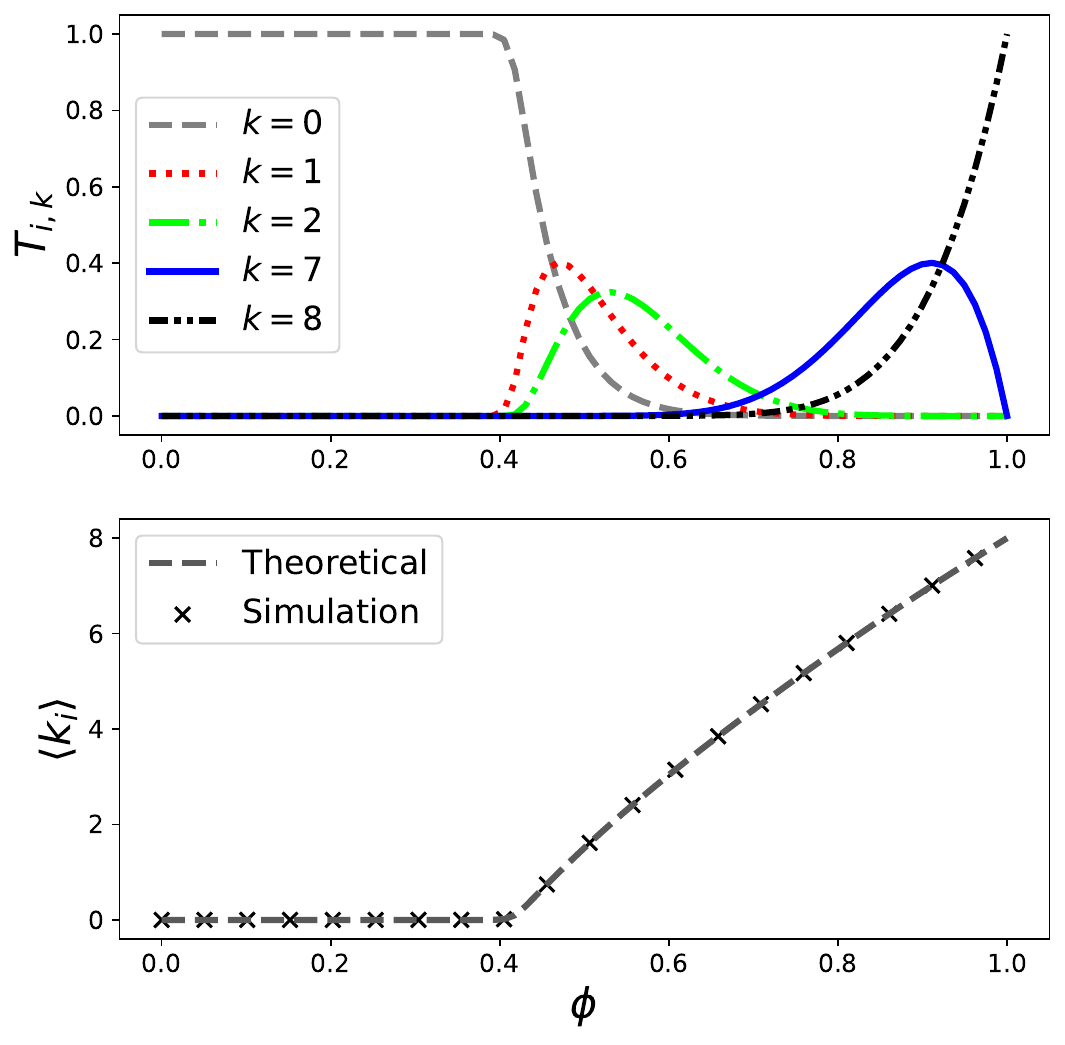}
\caption{(Top) Components of the degree distribution of occupied edges $T_{i,k}$ from Eq \ref{eq:Tki} for a degree 8 node in the GCC as a function of $\phi$. (Bottom) The average degree given by Eq \ref{eq:degdist} for the same node alongside simulation. The network is an Erd\H{o}s-Renyi graph with 10000 nodes and an overall average degree of 2.5.}
\label{fig:stacked_Tk_plot}
\end{figure}




\section{Applications}

In this section we will consider some examples of the reformulation that exhibit its advantage over the canonical approach.

\subsection{Sequential percolation}
 Let us define \textit{independent sequential percolation} (ISP) as the dynamical process of $n$ sequential and independent bond percolation processes over a network. At each pass of percolation, edges are occupied uniformly with probability $\phi_\ell\in\vec\phi=(\phi_1,\dots,\phi_n)$. The fraction of the network occupied by the largest component of nodes that are connected by maximally occupied edges can be calculated by passing the following messages over the edges of the graph
\begin{equation}
    \sigma_{i\leftarrow r}(z) = z\sum_{l=1}^{|\mathcal N(r)\backslash i|}\sum_{T\in W_l\backslash i}\left( \psi^{|T|}\prod_{t\in T}\sigma_{r\leftarrow t}\prod_{t'\in \bar T}[1-\sigma_{r\leftarrow t'}\phi_\ell]\right),\label{eq:ISP1}
\end{equation}
with $W_l$ taking its definition as before and where 
\begin{equation}
    \psi = \prod_{\ell=1}^n\phi_\ell.
\end{equation}
The size of the GCC is given by $ S(\vec  \phi) = \sum_i \sigma_i(1)/N$, with 
\begin{equation}
    \sigma_i(z)=z\sum_{l=1}^{|\mathcal N(i)|}\sum_{X\in Y_l}\left(\psi^{|X|}\prod_{r\in X}\sigma_{i\leftarrow r}\prod_{r'\in \bar X}[1-\sigma_{i\leftarrow r'}\phi_\ell]\right),\label{eq:ISP2}
\end{equation}

The fraction of nodes that were occupied by only a subset of the percolation processes can also be calculated by allowing only those $\phi_\ell$ pertaining to the subset in the product defining $\psi$. This is an immediate generalisation of Newman and Ferrario's coinfection model \cite{Newman_Ferrario} from two coinfecting diseases to $n$. By modulating $\psi$ we can also obtain quantities that are not available via their method, such as the fraction of the population that was coinfected by $n'<n$ diseases. We show an example of the ISP process in Fig \ref{fig:Tk_i_expr_stack} (top) for $n=2$; the circle markers track the simulation results of the primary percolation, whilst the triangle scatter points are the secondary process.
\begin{figure}[ht!]
\centering
\includegraphics[width=0.4\textwidth]{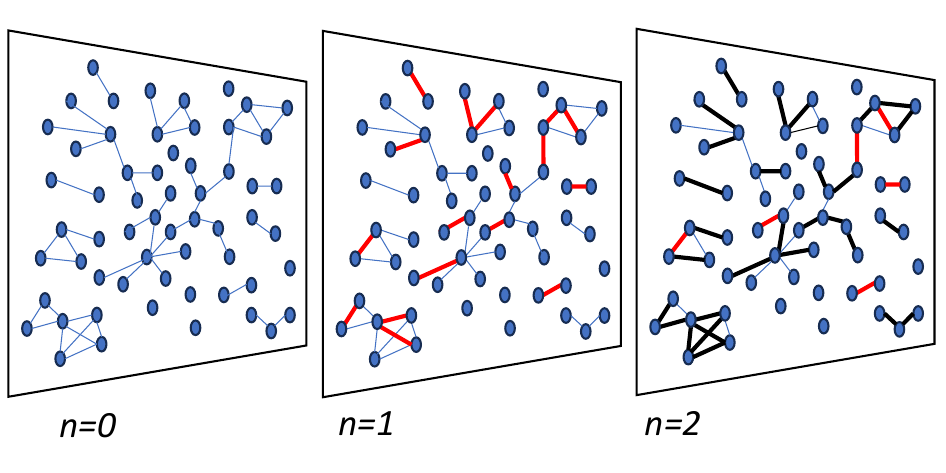}
\caption{A depiction of the sequential percolation process for $\phi_1<\phi_2$.}
\label{fig:seq}
\end{figure}

Treating the sequential processes as the composition of \textit{independent} dynamical processes is a simplification of the reality of many phenomenological process. Often, temporally separated processes are \textit{coupled} together in some manner. For instance, seasonal diseases can confer partial immunity to an individual, making them less likely to contract a subsequent strain \cite{PhysRevE.103.062308}; conversely, immunosuppressant effects could act to increase the likelihood of catching something else \cite{PhysRevE.103.042307, PhysRevE.106.014304}. To model this behaviour we must occupy edges with probabilities based on their occupation state of all of the previous percolations, which is analytically a harder task. Due to this, the $\ell$th percolation process in the sequence observes $2^{\ell-1}$ possible edge types and therefore requires this many parameters. For the case of $n=2$, we require 3 bond occupation parameters: $\phi_1$ for the first process; and $\phi_2$ and $\phi_3$ for the second process that occupies 1-occupied and 1-unoccupied edges, respectively. We can compute the GCC of maximally occupied edges following \textit{coupled sequential percolation} (CSP) by passing the following message sets over the edges
\begin{subequations}
\begin{align}
    \varphi_{i\leftarrow r} =& \sum_{l=1}^{|\mathcal N(r)\backslash i|}\sum_{T\in W_l\backslash i}\left(\prod_{t\in T}f_{r\leftarrow t}(\vec \phi)\prod_{t'\in \bar T}g_{r\leftarrow t'}(\vec \phi)\right)\big/\nu_{i\leftarrow r}\label{eq:coinf1},\\
    \omega_{i\leftarrow r} =& \prod_{t\in \mathcal N(r)\backslash i}g_{r\leftarrow t}(\vec \phi)\big/\mu_{i\leftarrow r}\label{eq:coinf2}
\end{align}
\end{subequations}
with 
\begin{align}
    f_{r\leftarrow t}(\vec \phi) =&\ \nu_{r\leftarrow t}\phi_1h_\varphi(\phi_2),\\
    h_y(\phi) =&\ 1-\phi + y_{r\leftarrow t}\phi,
\end{align}
and
\begin{align}
    g_{r\leftarrow t}(\vec \phi) =\ & \mu_{r\leftarrow t}(1-\phi_1) + (1-\mu_{r\leftarrow t})(1-\phi)h_\varphi(\phi_2)\nonumber\\
    &+ \mu_{r\leftarrow t}\phi_1h_\omega(\phi_3).
\end{align}
Expression \ref{eq:coinf1} accounts for nodes in the GCC of the first process and reduces to Eq \ref{eq:nujkmu} when $\phi_2=\phi_3=0$; whilst Eq \ref{eq:coinf2} accounts for nodes in the residual graph and reduces to Eq \ref{eq:mu_j} under the same limit. The size of the GCC formed by maximally occupied edges is given by 
\begin{equation}
    S = \frac 1 N \sum_i \nu_i - \frac 1N\sum_i\varphi_i,
\end{equation}
where $\nu_i$ is defined in Eq \ref{eq:nuij} and $\varphi_i$ is 
\begin{equation}
    \varphi_{i} = \sum_{l=1}^{|\mathcal N(i)|}\sum_{X\in Y_l}\left(\prod_{r\in X}f_{i\leftarrow r}(\vec \phi)\prod_{r'\in \bar X}g_{i\leftarrow r'}(\vec \phi)\right).
\end{equation}
This model is the 2-strain message passing equivalent of Eq 37 in \cite{Mann_Smith_Mitchell_Dobson_2022} re-written in the alternative formulation presented here. It can instantly be seen that the reformulated description presented here is far simpler.

\subsection{Non-binary percolation}
\begin{figure}[ht!]
\centering
\includegraphics[width=0.5\textwidth]{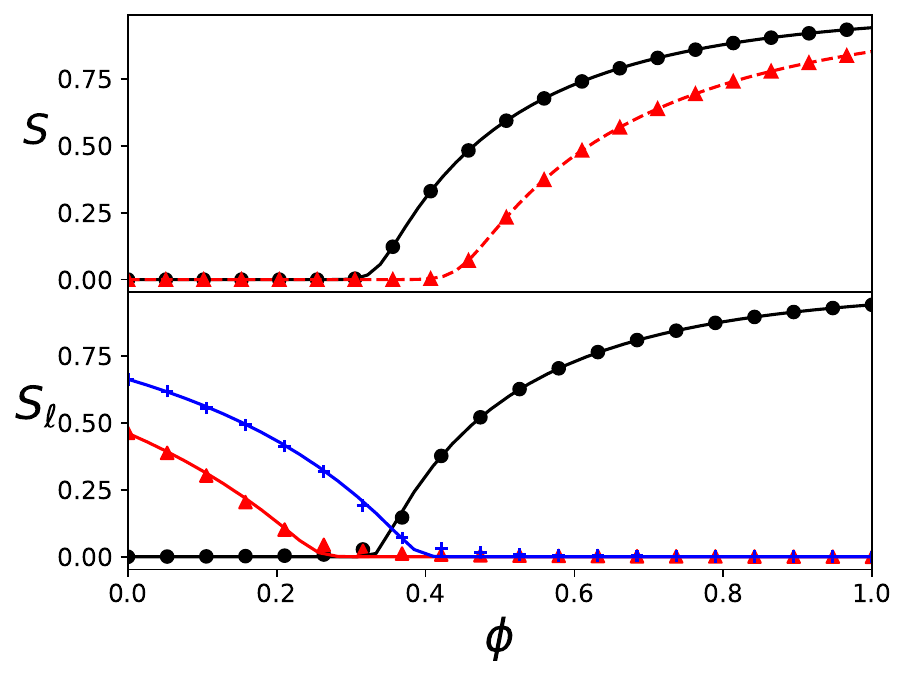}
\caption{Results for the ISP model (top) and the Potts-like model (bottom) on Erd\H{o}s-Renyi graphs with $10,000$ nodes. Scatter points are the average of 100 repetitions of Monte Carlo simulation whilst lines are the theoretical predictions of our models when passing Eqs \ref{eq:ISP2} and \ref{eq:pottslike}, respectively. The black circles are $\ell=1$, the red triangles are $\ell=2$ and the blue crosses are $\ell=3$.}
\label{fig:Tk_i_expr_stack}
\end{figure}

Let us now consider the application of our expression to a third kind of  percolation model. The equilibrium of ordinary bond percolation is a binary state fixed point containing occupied and unoccupied edge states. Let us next consider a Potts-like process \cite{Potts_1952, Wu_1982} over the edges of network in which edges can belong to one of $m$ states $q_1,\dots, q_m$ with occupation probabilities $\vec \phi =(\phi_1,\dots, \phi_m)$ with the constraint that $\sum_\ell\phi_\ell=1$ for $\ell\in [1,m]$. When $m=2$ the model reduces to ordinary bond percolation and $\vec\phi=(\phi, 1-\phi)$. For each edge-state we are concerned with finding the largest connected component that can be created by following edges of a particular state. Nodes may belong to the GCCs of multiple edge-states. For edge-state $q_\ell$ let $\nu_{\ell, i}$ be the probability that node $i$ with degree $|\mathcal N_i|$ belongs to the GCC of $q_\ell$-state occupied edges
\begin{equation}
    \nu_{\ell,i}(\vec \phi) = \sum_{\underset{X_{\ell}\neq\emptyset}{X_{1},\dots,X_{m}=\mathcal N(i)}}\left(
    \prod_{k=1}^m \prod_{r\in X_k} \nu_{k,i\leftarrow r}\phi_k\right),\label{eq:pottslike}
\end{equation}
with the condition $X_{\ell}\neq\emptyset$ indicating the node belongs to the $q_\ell$-state GCC. The summation is over all covering disjoint subsets (partitions) of the set of neighbours $\mathcal N(i)$ such that $X_\ell$ is not empty
\begin{equation}
    \left\{(X_1,\dots, X_m)\ \big|\ \ \bigcup_kX_k=\mathcal N(i),\ X_\ell \neq\emptyset\right\}
\end{equation}
A given covering set $X_1,\dots,X_m$ represents an edge configuration of the neighbours among the states. Index $k$ runs over the edge states $1,\dots,m$ and $r\in X_k$ is an index over the neighbours that are in state $q_k$ for the edge configuration. The message updates $\nu_{k,i\leftarrow r}$ are given by 
\begin{equation}
    \nu_{k,i\leftarrow r}(\vec \phi) =  \sum_{\underset{X_{k}\neq\emptyset}{X_{1},\dots,X_{m}=\mathcal N_r\backslash i}}\left(
    \prod_{j=1}^m \prod_{t\in X_j} \nu_{j,r\leftarrow t}\phi_j\right),
\end{equation}
 which is the probability that at least one of neighbour $r$'s neighbours (other than $i$) connects to the GCC of the $k$ edge-state. The size of the giant connected component for the $\ell$th state is given by
\begin{equation}
    S_\ell(\vec\phi) = \frac 1 N \sum_i \nu_{\ell,i}(\vec\phi).
\end{equation}
We show an example of this in Fig \ref{fig:Tk_i_expr_stack} (bottom) for $n=3$ with the following bond occupation probabilities for the three states: $\phi, \alpha(1-\phi)$ and $ 1 - \phi - \alpha(1-\phi)$ setting $\alpha=0.45$.

\section{Conclusion}


To conclude, we have introduced an alternative exact analytical formulation of the message passing algorithm on graphs. Our theoretical models provide additional insight into the environment of nodes embedded into the GCC beyond those that the traditional approach affords. We have shown that this approach lends itself to multiple extensions of message passing on graphs, in particular to systems such as sequential percolation and Potts-like models with multiple edge states. In both of those cases, the model presented here is simpler and more tractable than those presented previously and is arguably the natural description of nodes that are embedded in the GCC.

In future work we intend to explore higher-order networks using this framework, such as motif \cite{PhysRevE.107.054303}, neighbourhood \cite{Cantwell_Newman_2019}, hypergraph and simplicial \cite{PhysRevE.106.034319} models. In particular, the ability to pick apart different occupation probabilities, both \textit{along} and \textit{through} each higher-order structure, could yield important information on community structure, centrality metrics or optimal motif projection \cite{PhysRevE.105.044314, PhysRevE.107.054303}. Additionally, we would like to investigate how this representation could be used to describe complex contagions such as opinion dynamics and percolation models where multiple edge occupations are required to include a node in the GCC \cite{PhysRevE.105.034306}. We also intend to generalise the CSP model to include different kinds of coupling patterns between the percolation processes in the spirit of \cite{Funk_Jansen_2010, Mann_Smith_Mitchell_Dobson_2022, Mann_Smith_Mitchell_Dobson_2021}. Finally, this method can be used for gaining insight into the training of neural network models for a multitude of deep learning applications which we also intend to explore.

\bibliography{ref}

\end{document}